\documentclass[a4paper]{article}
\usepackage{Odyssey2022}
\usepackage{epsfig,amssymb,amsmath}
\usepackage{hyperref}
\usepackage{makecell}
\usepackage{multirow}
\usepackage[flushleft]{threeparttable}
\hypersetup{colorlinks=true}
\usepackage[table]{xcolor}
\ninept

\newcommand\mat[1]{\mathbf{#1}}
\newcommand\textgreen[1]{\textcolor[rgb]{0,0.5,0}{\textbf{#1}}}
\def\shade{\cellcolor{gray!20}}

\newcommand\Yang[1]{}
\newcommand\Quan[1]{}
\newcommand\Yiling[1]{}
\newcommand\Jason[1]{}
\newcommand\Ignacio[1]{}

\title{Attentive Temporal Pooling for Conformer-based \\
Streaming Language Identification in Long-form Speech}

\name{Quan Wang$^{*}$, Yang Yu$^{*}$, Jason Pelecanos, Yiling Huang, 
Ignacio Lopez Moreno\thanks{* Equal contribution.}}

\address{Google LLC\\
\small \tt {\{\href{mailto:quanw@google.com}{quanw},\href{mailto:yyuyy@google.com}{yyuyy},\href{mailto:pelecanos@google.com}{pelecanos},\href{mailto:yilinghuang@google.com}{yilinghuang},\href{mailto:elnota@google.com}{elnota}\}@google.com}
}
\begin{document}
\maketitle

\begin{abstract}
In this paper, we introduce a novel language identification system based on conformer layers. We propose an attentive temporal pooling mechanism to allow the model to carry information in long-form audio via a recurrent form, such that the inference can be performed in a streaming fashion. Additionally, we investigate two domain adaptation approaches to allow adapting an existing language identification model without retraining the model parameters for a new domain. We perform a comparative study of different model topologies under different constraints of model size, and find that conformer-based models significantly outperform LSTM and transformer based models. Our experiments also show that attentive temporal pooling and domain adaptation improve model accuracy.
\end{abstract}

\section{Introduction}
\label{sec:intro}

Language Identification (LangID / LID) is the task of automatically identifying the spoken language of a digitized speech utterance~\cite{zissman1996comparison,lopez2014automatic,ambikairajah2011language,zissman2001automatic}, which has been widely used in various modern speech processing systems. Common applications include automatic call routing, multilingual speech transcription, multilingual speech translation, and content-based audio retrieval~\cite{muthusamy1994reviewing,ambikairajah2011language}.
Particularly, a commercially viable application in recent years is the interactive voice assistive system, where LangID can be used to automatically select the output from several automatic speech recognition (ASR) models that run in parallel, such that the user is able to interact with the system with multiple different languages
~\cite{johan2018teaching,wan2019tuplemax}.
Additionally, there is  an  increasing  research  interest  in  multilingual ASR systems that are suitable for code-switching applications. For such applications, LangID can either be jointly trained with the ASR model~\cite{waters2019leveraging,punjabi2021joint,duroselle2021modeling}, or provide auxiliary input to the ASR model~\cite{lyu2008language,zeng2018end,li2019towards}.

In this paper, we are particularly interested in applications where language identification is used for analyzing streaming long-form audio, such as podcast, telephony speech, and video streaming. Language identification results produced at runtime can be used for various purposes, such as triggering ASR captioning with the right language, context-based indexing and searching, triaging the content for human auditors, recommending the content to the most relevant audience, and delivering advertisement in the right language.
Such applications usually have these requirements:

\begin{enumerate}
    \item \textit{Low latency}. In streaming applications, it is critical to produce the language prediction signal accurately in a timely manner to be consumed by downstream components such as ASR and natural language understanding (NLU). Thus some non-causal model topologies, such as bi-directional LSTM, can be detrimental.
    \item \textit{No recency bias}. Long-form speech often allows the model to predict the language based on a relatively long context. However, recurrent neural networks such as LSTM~\cite{hochreiter1997long} often suffer from a bias towards short-term context~\cite{gonzalez2014automatic}.
    \item \textit{Noise robustness}. Long-form utterances are often interleaved with silence, non-speech, as well as speech segments from various sources, including different speakers, difference devices, different reverberant conditions, and different signal-to-noise ratio (SNR). The non-speech segments and noisy speech segments may cause degradation of the overall language prediction accuracy if we simply average the results from all segments.
    \item \textit{Parallelization}.  As deep learning accelerators become available in more and more cloud computing services~\cite{tpuv4blog} as well as consumer devices~\cite{googletensorblog}, neural networks that can perform inference in parallel will better benefit from the computational power of such hardware.
    In many scenarios, such as high-load services and on-device applications, parallelizable models are largely preferred over non-parallelizable models.
\end{enumerate}

Based on these requirements, we introduce a language identification model based on conformer layers~\cite{gulati2020conformer} and an attentive temporal pooling mechanism\footnote{An open source implementation of the attentive temporal pooling module based on Lingvo~\cite{shen2019lingvo} is provided at: \url{https://github.com/google/speaker-id/tree/master/lingvo}}. The inference of this model runs in a streaming fashion, and the conformers can be easily parallelized on accelerator hardware to minimize the latency. The attentive temporal pooling mechanism computes an attention weight for each temporal step of the conformer model, and uses a weighted moving average to produce an aggregated embedding at each step. This mechanism helps the model to better attend to speech segments that contain more information related to the spoken language.


At the same time, different application domains may have (i) different prior distributions of languages and (ii) different data properties. Here we study two simple domain adaptation approaches that can improve the empirical performance of the LangID model in various application domains without retraining the neural network. This significantly reduces the development cost of LangID models while achieving improved performance for each application.

The rest of this paper is organized as follows. In Section~\ref{sec:related}, we will briefly review existing works that are related to our LangID system. In Section~\ref{sec:sys}, we will introduce our LangID system in detail, including the feature frontend in Section~\ref{sec:frontend}, the data augmentation strategy in Section~\ref{sec:aug}, the model topology in Section~\ref{sec:topology}, the attentive temporal pooling mechanism in Section~\ref{sec:pooling}, and the domain adaptation methods in Section~\ref{sec:adapt}.
We describe our experiments and results in Section~\ref{sec:exp}, and present our conclusions in Section~\ref{sec:conclusion}.

\section{Related Work}
\label{sec:related}

LangID is also often referred to as Spoken Language Recognition (SLR)~\cite{li2013spoken,snyder2018spoken,valk2021voxlingua107} to avoid confusion with text-based language identification systems~\cite{jauhiainen2019automatic}.
While multiple systems have been proposed to exploit the acoustic, phonetic, morphologic and semantic level representations of the utterances ~\cite{wang2019signal,chandak2020streaming,jauhiainen2019automatic}, most common LangID systems today directly take acoustic features of the utterance as input, such as Mel-frequency cepstral coefficients (MFCC) or log Mel-filterbank energies (LFBE). 

Since the proposal of one of the earliest HMM-based LangID systems in 1977~\cite{house1977toward}, various models and approaches have been explored. Unlike speech recognition and speech synthesis, where the task can be viewed as a \textit{``sequence transduction''} problem, LangID is usually considered as a \textit{``sequence summation''} problem. Thus many LangID systems are largely inspired by speaker recognition systems. For example, i-vector~\cite{dehak2010front} based LangID systems had been very popular in early 2010s~\cite{dehak2011language,martinez2011language}. 

As deep learning has gained in popularity, most modern LangID systems are based on neural networks. Deep feed-forward neural network (DNN) based LangID models had been proven significantly more accurate than i-vector based models~\cite{lopez2014automatic,snyder2018spoken}. Convolutional neural networks (CNN) and recurrent neural networks (RNN) such as LSTM~\cite{hochreiter1997long} had also been explored to reduce model sizes as well as improve LangID accuracy~\cite{lei2014application,gonzalez2014automatic}.

Conformers are convolution-augmented transformers that are used as building blocks in many different deep learning systems~\cite{gulati2020conformer}. It was originally proposed for speech recognition, but has found application in many other tasks such as speech enhancement~\cite{o2021conformer,narayanan2021cross}, speech separation~\cite{guo2021recent}, speaker diarization~\cite{liu2021end}, and sound event detection~\cite{miyazaki2020conformer}. In ~\cite{duroselle2021modeling}, a conformer-based feature extractor is used for joint ASR and LangID training. In ~\cite{lyu2021ant,wangroyalflush},
conformer-based models are first pretrained for ASR tasks, then used for transfer learning on various language recognition tasks~\cite{wang2021olr}.
We chose conformers as the basic building block of our LangID system due to its organic combination of convolution and multi-head self-attention, the demonstrated accuracy improvements, and its ability to parallelize on accelerator hardware.

Various works have examined statistics pooling~\cite{salmanjournal2011, salman2012thesis, snyder2018spoken} and attention modeling~\cite{zhu2018self,chowdhury2018icassp} for speaker recognition,
where the focus is mostly on the utterance level analysis. However, for real-time systems where latency is critical, there is a need to provide intermediate inference outputs in an online fashion. In this paper, we share a straightforward approach for generating these statistics using either segment level or per-frame recursion analysis for LangID.

\section{System Description}
\label{sec:sys}

\subsection{Feature frontend}
\label{sec:frontend}

In the feature frontend of our LangID system, we first apply automatic gain control~\cite{prabhavalkar2015automatic} to the input audio, then extract 32ms Hanning-windowed frames with a step of 10ms. For each frame, 128-dimensional log Mel-filterbank energies (LFBE) are computed in the range between 125Hz and 7500Hz. These filterbank energies are then stacked by 4 frames and subsampled by 3 frames, resulting in final features of 512 dimensions with a frame rate of 30ms.

\subsection{Data augmentation}
\label{sec:aug}

To make sure our LangID model is robust against various acoustic environments and noise conditions, we apply data augmentation during the training of the LangID model.
We randomly apply multi-style training (MTR)~\cite{lippmann1987multi,ko2017study} to part of the training utterances with an SNR ranging from 5dB to 25dB. The noise source consists of ambient noises recorded in cafes, kitchens, vehicles, and quiet environments, as well as audio clips of music and sound effects downloaded from the YouTube Audio Library\footnote{\url{https://youtube.com/audiolibrary}} and Getty Images\footnote{\url{https://www.gettyimages.com/about-music}}. The room configurations consist of 24 million convolutional room impulse responses generated by a room simulator~\cite{kim2017generation}. 

We also apply SpecAugment~\cite{park2019specaugment} to the rest of the training utterances where MTR did not apply. We found this practice pretty useful, as applying both MTR and SpecAugment to the same utterance can easily lead to over-augmentation.

\subsection{Model topology}
\label{sec:topology}

\begin{figure}[t!]
      \includegraphics[width=0.49\textwidth]{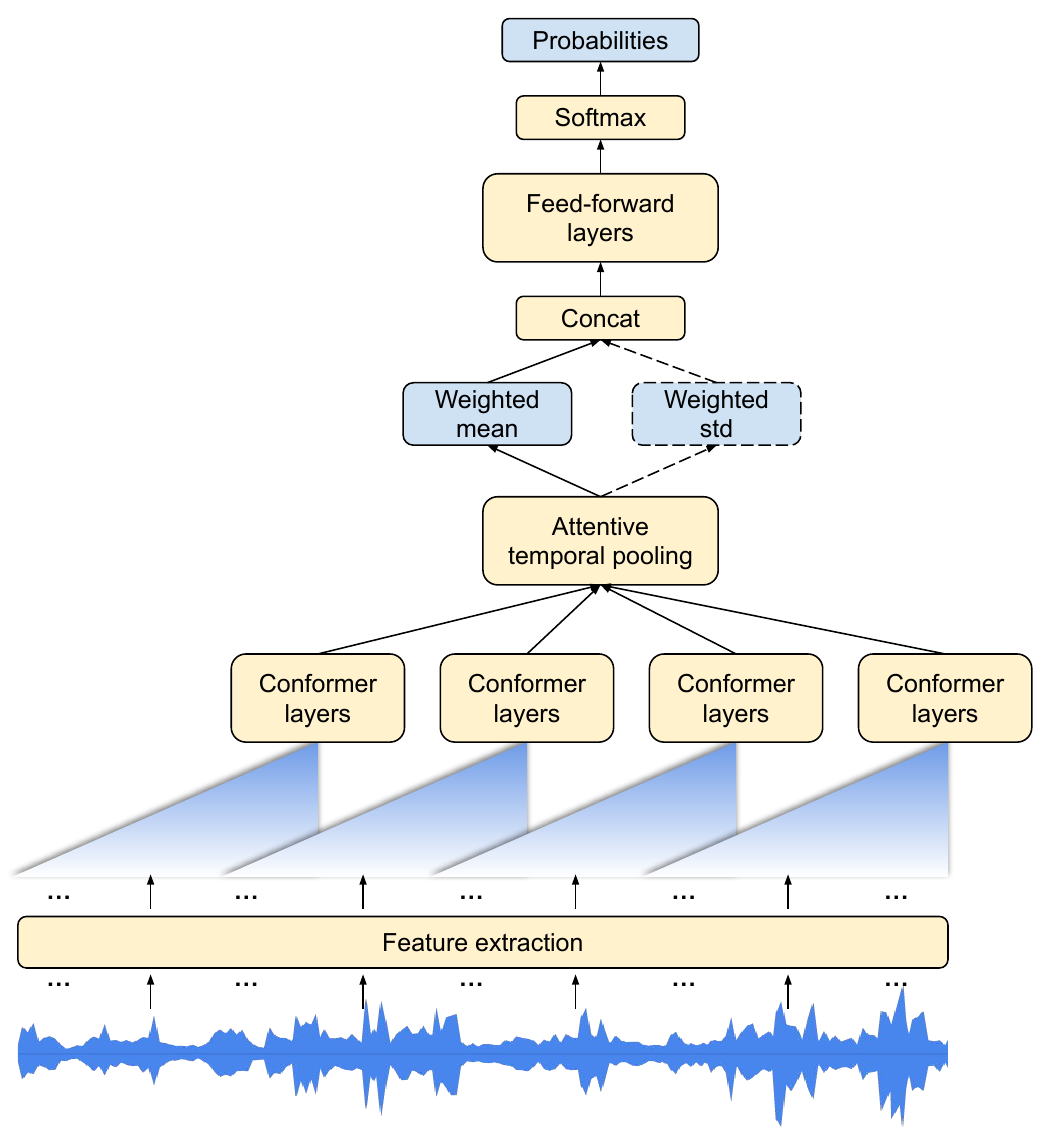}
     \caption{\it Diagram of our LangID model topology. The shaded triangles represent the receptive fields of the conformer layers.}
     \label{fig:diagram}
\end{figure}

The diagram of our conformer-based LangID model topology is shown in Fig.~\ref{fig:diagram}.
After performing data augmentation followed by feature extraction, we
add absolute positional encodings~\cite{vaswani2017attention} to these features, and feed them to the conformer encoder, which has a stack of 12 conformer layers. Similar to the original conformer paper~\cite{gulati2020conformer}, we explore three different model sizes, where the dimensionality of each layer is 144 (small), 256 (medium), and 512 (large), respectively. Each layer has a multi-head self attention with 8 heads. The 1-D depth-wise convolutional components span 32 elements. Additionally, we perform a stack-by-2 then subsample-by-2 operation on the third conformer layer output to reduce inference cost, and insert a non-linear projection layer with output dimension of 144/256/512 (for S/M/L, respectively) after the fourth conformer layer. This means each inference step would be of size 2 (every 2 frames), covering a receptive field of about 0.06 seconds of the input. This conformer encoder generates a 144/256/512 (for S/M/L, respectively) dimensional output for each input frame. The forward propagation calls of individual layers within the conformer network for different temporal positions are independent of each other. Thus they can be batched to run in parallel on accelerator hardware during both training and inference time.

The outputs of these conformer layers will be sent to the attentive temporal pooling layer to produce a weighted mean vector, which will be optionally concatenated with a weighted standard deviation vector. Then this weighted vector will be fed into a feed-forward network, consisting of a 256-dim layer with ReLU activation, and a linear projection whose output dimension equals the number of candidate languages. Finally, a softmax layer is applied to produce the probability distribution of different language candidates.

\subsection{Attentive temporal pooling}
\label{sec:pooling}

We introduce a weighted moving average implementation to allow the attentive temporal pooling mechanism to run in an online fashion.
Assume that at time step $t$, the conformer layers produce an output embedding $\mat{h}_t$. The attentive temporal pooling module will compute an attention weight $w_t$ based on the current embedding:
\begin{equation}
    w_t = f_\text{att}(\mat{h}_t) + \epsilon .
\end{equation}
The function $f_\text{att}(\mat{h}_t)$ is calculated as the sigmoid activation of a linear transform of $\mat{h}_t$. A small value ($\epsilon=0.0001$) is added to ensure numerical stability. 

The following sufficient statistics (counts, sums and sums of squares) are tracked at time $t$:
\begin{eqnarray}
    \eta_t &=& \sum_{s=1}^{t} w_s ,\\
    \boldsymbol{A}_t &=& \sum_{s=1}^{t} w_s \mat{h}_s ,\\
    \boldsymbol{Q}_t &=& \sum_{s=1}^{t} w_s \mat{h}_s^2.
\end{eqnarray}
Note here and elsewhere that $(\cdot)^2$ is used to denote the element-wise square.

Then the weighted mean $\boldsymbol{\mu}_t $ and weighted standard deviation $\boldsymbol{\sigma}_t$ can be calculated only using the  sufficient statistics at time $t$:
\begin{eqnarray}
    \label{eq:weighted_mean}
    \boldsymbol{\mu}_t &=& \frac{\boldsymbol{A}_t}{\eta_t} ,\\
    \label{eq:weighted_std}
    \boldsymbol{\sigma}_t &=& \sqrt{\frac{\boldsymbol{Q}_t}{\eta_t} - \boldsymbol{\mu}_t^2} .
\end{eqnarray}

To build the attentive temporal pooling as part of our neural network inference graph, and to handle streaming inference, we utilize the following \textbf{recurrent form} of the sufficient statistics in our implementation:
\begin{eqnarray}
    \eta_t &=& \eta_{t-1} + w_t ,\\
    \boldsymbol{A}_t &=& \boldsymbol{A}_{t-1} + w_t \mat{h}_t ,\\
    \boldsymbol{Q}_t &=& \boldsymbol{Q}_{t-1} + w_t \mat{h}_t^2 .
\end{eqnarray}
Note that they can be initialized as $\eta_0=0$, $\boldsymbol{A}_0 = \boldsymbol{0}$ and $\boldsymbol{Q}_0 = \boldsymbol{0}$. This representation allows us to incrementally calculate $\{\boldsymbol{\mu}_t, \boldsymbol{\sigma}_t\}$ from the previous sufficient statistics (state variables in the inference graph) $\{\eta_{t-1}, \boldsymbol{A}_{t-1}, \boldsymbol{Q}_{t-1}\}$ and the current intermediate network output frame $\mat{h}_t$. The streaming model in this paper uses this recurrent formulation interpretation.

This attentive temporal pooling mechanism can be easily extended to a multi-head version, where each output embedding $\mat{h}_t$ produces multiple weights, and we concatenate multiple weighted mean and weighted standard deviation vectors as the input to the feed-forward layers. However, in our LangID experiments, multi-head attentive temporal pooling has very similar performance with single-head attentive temporal pooling.

\subsection{Domain adaptation}
\label{sec:adapt}

\begin{figure}[t!]
\begin{center}
      \includegraphics[width=0.35\textwidth]{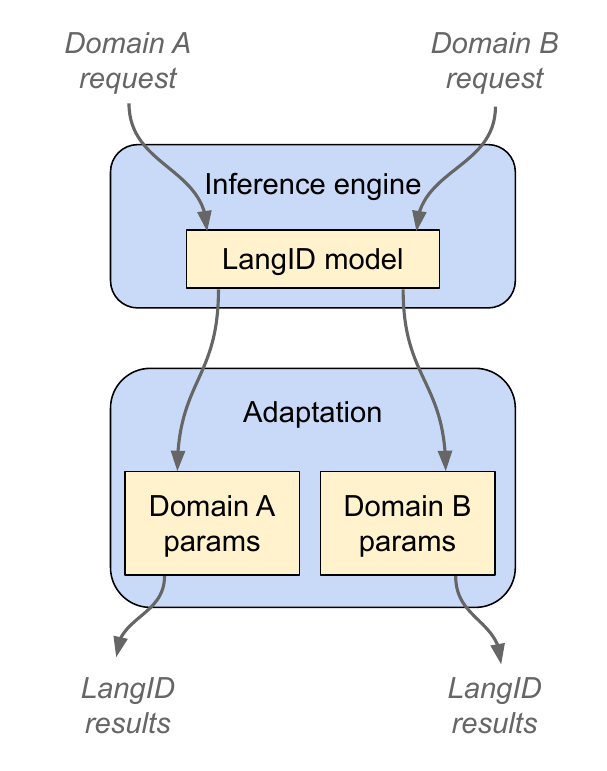}
     \caption{\it Domain adaptation allows us to deploy a unified inference backend that handles requests from different domains with a single LangID neural network model. The inference results are post-processed with adaptation parameters that are dynamically selected based on the domain information in the request.}
     \label{fig:adaptation}
     \end{center}
\end{figure}

Training a neural network model for the LangID system described above can be very expensive, especially when the model is trained on massively multilingual datasets. For simplicity, when we deploy the LangID system to different applications, it is usually desired that the same neural network model is deployed to all applications, while each application has the ability to adjust the output probabilities to make it consistent with the application specific class prior probabilities and perhaps data differences.

For example, assume we want to deploy a LangID model to identify the language in videos for two different video hosting websites. It is possible that most videos on website A are in English, while most videos on website B are in Spanish. At training time, we could include datasets from both websites to make our model robust. However, when deploying the model, we want the system to incorporate such prior information and additional data related differences. 

One approach is the use of a Gaussian backend classifier applied to i-vectors~\cite{mccree2014_1} or x-vectors~\cite{snyder2018spoken}. In this work, we explore basic approaches of what can be done if only the model output probabilities are available. For example, the results are generated by a cloud service. Here we describe two domain adaptation approaches: \textit{prior replacement} and a discriminatively trained \textit{output transform}. They do not require updating the LangID neural network itself, thus making it possible to deploy a unified inference backend with one LangID model to handle requests from different domains, as shown in Fig.~\ref{fig:adaptation}.

\subsubsection{Prior replacement}

In the first approach, we re-estimate the posterior probabilities of the languages for a specific application, using the new class prior probabilities given the new target domain data. Suppose we have a set of randomly drawn samples from the target application data. We then count the number of samples drawn for each of the $K$ languages $\{L_i\}_{i=1,\ldots,K}$. Let $c_i$ be the number of samples related to language $L_i$. A smoothed estimate of the class prior probabilities may be given as~\cite{baker2004_1, chen1998_1}:

\begin{equation}
\label{eq:class_prior_prob_estimate}
  P(L_i|D_{new}) = \frac{c_i + R}{\sum_{j=1}^K (c_j + R)} ,
\end{equation}
where $R$ is the prior data relevance count. (Empirically, when the sample size is $N=10000$, we found $R=4$ to work well.)

We treat the softmax outputs from the LangID neural network as the probabilities of the languages given the input utterance. The LangID neural network was trained with equal class prior probabilities. Given the new application specific prior probabilities, and a neural network trained on uniform class prior probabilities, the posterior probability of a language given the new application target and utterance $X$ is computed as~\cite{bailerjones2011combining}:
\begin{equation}
\label{eq:application_posterior_probability}
  P(L_i|X, D_{new}) = \frac{P(L_i|D_{new})P(L_i|X,D_{old})}{\sum_{j=1}^K P(L_j|D_{new})P(L_j|X,D_{old})} .
\end{equation}
Here, $D_{old}$ represents the information regarding class priors built into the existing model trained on the old data. In this work, the language class priors are uniform and this gives a simplified version of the result in~\cite{bailerjones2011combining} because the common values cancel. $D_{new}$ represents information regarding the class priors for the new application. The term $P(L_i|D_{new})$ is the new estimate of the application specific class prior probabilities, and $P(L_i|X,D_{old})$ is the probability generated by the softmax of the neural network based on the old equal class prior assumptions.

\subsubsection{Output transform}

In the second approach, we assume a dev-set from the new target domain is available, and we optimize a transform on this dev-set.

Assume that the original output of the LangID model is denoted as a probability distribution over $K$ languages $\mat{p}\in [0,1]^K$. For the new domain, let $\mat{a}$ and $\mat{b}$ be two $K$-dimensional vectors.
We transform $\mat{p}$ into a new probability distribution:
\begin{equation}
\label{eq:adapt}
     \widetilde{\mat{p}}=\text{softmax}(\mat{a} \odot \log \mat{p} + \mat{b}) ,
\end{equation}
where $\odot$ denotes element-wise multiplication.

Given a dev-set of $N$ samples from this specific domain ($N=10000$ in our experiments), we optimize $\{\mat{a},\mat{b}\}$ to minimize a regularized cross entropy between $\widetilde{\mat{p}}$ and the ground truth language $\mat{y}$ on this dev-set:
\begin{equation}
    \label{eq:adapt_optimize}
    \underset{\mat{a},\mat{b}}{\arg \min} \Big( \frac{1}{N} \sum_{i=1}^N L_\text{cent}(\widetilde{\mat{p}}^{(i)},\mat{y}^{(i)}) + w_\text{reg}(||\mat{a}-\mat{1}|| + ||\mat{b}||)\Big),
\end{equation}
where $L_\text{cent}$ denotes cross entropy loss and $w_\text{reg}$ is the weight for the regularization term.

Once we have optimized $\{\mat{a},\mat{b}\}$ on the dev-set, we can deploy Eq.\ref{eq:adapt} as a post-processing step together with the LangID model without modifying the model itself as shown in Fig.~\ref{fig:adaptation}. At the same time, while the LangID model is trained on batched short segments for efficiency, Eq.~\ref{eq:adapt_optimize} is based on long-form inference outputs. This can better compensate for the duration gap between training and inference.


Although Eq.~\ref{eq:adapt} can be replaced by more complicated forms, such as using a full $K \times K$ transform matrix instead of the vector $\mat{a}$, or even using a tiny neural work, such approaches can easily lead to overfitting on the dev-set, especially that the dev-set is usually much smaller than the training set. We found that adaptation with only $2K$ parameters (\emph{i.e.} $\mat{a}$ and $\mat{b}$) effectively improves in-domain performance without overfitting the dev-set.

Interestingly, if $\mat{a}$ is fixed as a vector of ones, the optimization result is closely related to the prior replacement method in Eq.~\ref{eq:application_posterior_probability}. The key differences are that parameters are trained discriminatively instead of directly using the class priors from the dev-set, and the regularization/constraints are different.


\section{Experiments}
\label{sec:exp}

\begin{table*}[t]
\caption{\label{tbl:model_comparison} {\it Comparison of the Conformer model with the LSTM and Transformer models. The size in MB is for models quantized to int8 and serialized to TFLite format~\cite{alvarez2016efficient,shangguan2019optimizing}. FLOP/s represents the number of floating point operations needed to process 1 second of audio.}}
\vspace{2mm}
\centerline{
\begin{threeparttable}
\begin{tabular}{c|c|c|c|c|cc}
\hline
\makecell{Model size} & \makecell{Encoder type} & \makecell{Number of \\layers} & \makecell{Layer \\dimensions} & \makecell{GFLOP/s} & \makecell{Voice query \\ avg. accuracy} & \makecell{Long-form \\ avg. accuracy} \\
\hline
\multirow{3}{*}{Small ($\sim$ 7MB)}
& LSTM & 4 & 1024 $\rightarrow$ 256$^\dagger$ & 0.46 & 77.55\% & 72.05\% \\
& Transformer & 14 & 144 & 0.76 & 79.86\% & 75.39\% \\
& Conformer & 12 & 144 & 0.45 & \textgreen{84.13\%} & \textgreen{75.60\%} \\
\hline
\multirow{3}{*}{Medium ($\sim$ 30MB)}
& LSTM & 4 & 2048 $\rightarrow$ 512$^\dagger$ & 1.58 & 84.66\% & 76.59\% \\
& Transformer & 14 & 256 & 5.37 & 86.99\% & 76.98\% \\
& Conformer & 12 & 256 & 1.91 & \textgreen{88.24\%} & \textgreen{77.26\%} \\
\hline
\multirow{3}{*}{Large ($\sim$ 120MB)}
& LSTM & 8 & 4096 $\rightarrow$ 1024$^\dagger$ & 11.09 & 85.72\% & 78.83\% \\
& Transformer & 14 & 1024 & 38.60 & 88.41\% & 79.09\% \\
& Conformer & 12 & 512 & 7.56 & \textgreen{89.58\%} & \textgreen{79.22\%} \\
\hline
\end{tabular}
\begin{tablenotes}
\item[$\dagger$] {\it The LSTM models have a pyramid structure~\cite{gonzalez2014real} with decreasing layer dimensions from bottom to top layers.}
\end{tablenotes}
\end{threeparttable}}
\end{table*}


\subsection{Datasets and evaluation metrics}
\label{sec:dataset_metric}

Our model is trained to distinguish between 65 different languages\footnote{The list of the 65 languages: Afrikaans, Amharic, Arabic, Azeri, Belarusian, Bulgarian, Bengali, Catalan, Chinese, Czech, Danish, German, Greek, English, Spanish, Basque, Farsi, Finnish, Filipino, French, Galician, Gujarati, Hebrew, Hindi, Hungarian, Armenian, Indonesian, Icelandic, Italian, Japanese, Javanese, Georgian, Khmer, Kannada, Korean, Lao, Lithuanian, Latvian, Malayalam, Marathi, Malay, Burmese, Norwegian, Nepali, Dutch, Polish, Portuguese, Romanian, Russian, Sinhala, Slovak, Slovenian, Serbian, Sundanese, Swedish, Swahili, Tamil, Telugu, Thai, Turkish, Ukrainian, Urdu, Vietnamese, Cantonese, Zulu.}. The training and evaluation utterances comprise anonymized voice queries from Google Assistant, and long-form utterances extracted from YouTube videos\footnote{The YouTube dataset only covers 61 languages, with Chinese, Filipino, Cantonese and Zulu missing.} transcribed by human annotators. The average length of a voice query is about 3.3 seconds with a standard deviation of 1.5 seconds,  while the average length of a long-form utterance is about 20.7 minutes with a standard deviation of 10.6 minutes. The size of the training set for each language varies from 1M to 20M utterances (including both voice queries and long-form); while the size of the evaluation set for each language is around 20K utterances for voice queries, and 20K utterances for long-form. We use a pretrained Voice Activity Detector (VAD) to remove the non-speech parts from the utterances for both training and evaluation. MTR and SpecAug are applied to the training utterances as described in Section~\ref{sec:aug}.  In this paper, we are interested in unconstrained language identification and use the softmax cross entropy loss with Adam optimization~\cite{kingma2014adam} in training\footnote{In applications where each request is constrained to a subset of candidate languages, the TupleMax loss~\cite{wan2019tuplemax} is preferred.}.

For each evaluation, we report the average accuracy of 65 languages as the model performance metric. For each language, the accuracy is defined by the percentage of the utterances whose ground truth language has the highest score in the predicted probability distribution. From a binary classification perspective, if we only accept the language when this language has the highest score in the predicted probability distribution, then this accuracy number is equivalent to the recall rate of this language.

\subsection{Comparison of encoder models}
\label{sec:conformer_model}

In our first experiment, we compare the performance of the LSTM, transformer and conformer architectures, which are the commonly used encoder models for speech recognition. As mentioned in Section~\ref{sec:topology}, we experiment with three different model size constraints: small, medium and large, where the number of parameters are around 7M, 30M, and 120M, respectively. For this experiment, no temporal pooling layers are included for all the models --- we directly take the last-frame embedding as the final embedding for the softmax layer. Configurations for each model architecture under each constraint are detailed as below.

\textbf{LSTM model}: The LSTM models we used in this experiment have similar topologies to the ones used in ~\cite{gonzalez2014real}. The model has a stack of LSTM layers, with each LSTM layer, except the last one, followed by a projection layer~\cite{sak2014long}. The LSTM layers have a pyramid-like shape, where the bottom layer is the largest one and the following layer dimensions decrease linearly. Experimentally, we find that both adding projection layers and reducing the size of layers further up in the network significantly speed up training and inference without hurting performance.

\textbf{Transformer model}: The transformer model is implemented based on the work of Transformer-XL~\cite{dai2019transformer}. The transformer models for all sizes have 14 transformer layers as we found that the model depth is important for model performance. The layer dimensions of small, medium, and large models are 144, 256, and 1024, respectively.

\Yiling{Do we need to explicitly mention that this is implemented with a wrapper on top of the Keras module, while the Conformer is using the Lingvo open-sourced library? This potentially implies that there could be some "unexplainable diffs" in this Transformer vs Conformer comparison...}

\textbf{Conformer model}: As described in Section~\ref{sec:topology}, the model has 12 conformer layers. The layer dimensions of small, medium, and large models are 144, 256, and 512, respectively, following the experiment setting from the original conformer paper~\cite{gulati2020conformer}.

The experimental results are shown in Table~\ref{tbl:model_comparison}. As we can see, under all different model size constraints, the conformer model shows the best performance, followed by the transformer model. This is consistent with observations from speech recognition experiments~\cite{gulati2020conformer}.
By counting \texttt{total\_float\_ops} with TensorFlow Profiler\footnote{\url{https://github.com/tensorflow/profiler}},
we also report the number of floating point operations needed to process 1 second of audio (FLOP/s) for each model in the table. Conformer models are relatively computationally efficient across the three model sizes. Given that conformer models are also more accelerator-friendly, they are the preferred choice from both performance and efficiency perspectives. 

\begin{table}
\caption{\label{tbl:pooling} {\it Comparison of different temporal pooling approaches based on the medium-size conformer model.  }}
\vspace{2mm}
\centerline{
\begin{tabular}{l|cc}
\hline
\makecell{Model} & \makecell{Voice query \\ avg. accu.} & \makecell{Long-form \\ avg. accu.} \\
\hline
Medium-size conformer & 88.24\% & 77.26\% \\
\rotatebox[origin=c]{180}{$\Lsh$} + Mean pooling & 88.18\% & 77.28\% \\
\rotatebox[origin=c]{180}{$\Lsh$} + Mean\&std pooling & 88.32\% & 77.35\% \\
\rotatebox[origin=c]{180}{$\Lsh$} + Weighted mean pooling & \textgreen{88.92\%} & 77.45\% \\
\rotatebox[origin=c]{180}{$\Lsh$} + Weighted mean\&std pooling & 88.74\% & \textgreen{77.81\%} \\
\hline
\end{tabular}}
\end{table}

\begin{table*}
\caption{\label{tbl:adapt} {\it Language identification total accuracy for different domain adaptation approaches. The model is a medium-size conformer with weighted mean pooling. Note that here we do not use ``average accuracy" as it assumes uniform prior distribution of all languages.}}
\vspace{2mm}
\centerline{
\begin{tabular}{c|c|c|c}
\hline
\makecell{Method} & \makecell{Domain  adapted to} & \makecell{Voice query  total accuracy \\ (Perplexity: $PP=33.8$)} & \makecell{Long-form  total accuracy \\ (Perplexity: $PP=56.2$)} \\
\hline
No adaptation & - & 90.85\% & 77.94\% \\ \hline
\multirow{2}{*}{Prior replacement ($R=0$)} & Voice query & \textgreen{91.76\%} & \shade{54.21\%} \\
 & Long-form & \shade{83.09\%} & \textgreen{78.18\%} \\ \hline
 \multirow{2}{*}{Prior replacement ($R=4$)} & Voice query & \textgreen{91.74\%} & \shade{74.67\%} \\
 & Long-form & \shade{90.16\%} & \textgreen{78.17\%} \\ \hline
\multirow{2}{*}{Output transform} & Voice query & \textgreen{92.32\%} & \shade{76.18\%} \\
& Long-form & \shade{76.97\%} & \textgreen{82.55\%} \\
\hline
\end{tabular}}
\end{table*}

\subsection{Attentive temporal pooling}
\label{sec:pooling_exp}

To evaluate the impact of the attentive temporal pooling described in Section~\ref{sec:pooling}, we trained four additional medium-size conformer models:
\begin{enumerate}
    \item Naive mean pooling, with equal weight for each frame;
    \item Naive mean and standard deviation pooling, with equal weight for each frame;
    \item Weighted mean pooling (Eq.~\ref{eq:weighted_mean}); 
    \item Weighted mean (Eq.~\ref{eq:weighted_mean}) and standard deviation (Eq.~\ref{eq:weighted_std}) pooling.
\end{enumerate}

The experimental results are shown in Table~\ref{tbl:pooling}. As we can see from the table, attentive temporal pooling shows improved language identification accuracy compared with no temporal pooling as well as the non-attentive naive pooling approaches.


\subsection{Domain adaptation}
\label{sec:adapt_exp}

As mentioned in Section~\ref{sec:dataset_metric}, our training and evaluation data comprise both anonymized voice queries and long-form speech from YouTube. These are two domains that are very different in many respects, including: (1) the textual content; (2) the length of the speech; and (3) the prior language distribution. At training time, we always train the LangID model on the joint dataset to increase model robustness and reduce development cost. But to achieve improved in-domain performance, we adapt our model to these two different domains with a held out dev-set from each domain, as described in Section~\ref{sec:adapt}.
The models with and without domain adaptation are then evaluated for each domain.

The evaluation results for domain adaptation are shown in Table~\ref{tbl:adapt}. Note that for this experiment, we report the ``total accuracy" for the entire evaluation dataset, instead of the ``average accuracy" over all languages. The latter asserts a uniform class prior distribution for all languages and may not reflect specifics of the application. As we can see, the two domain adaptation methods --- prior replacement and the discriminatively trained output transform --- improve the language identification total accuracy on both domains. 

The prior replacement approach shows a larger improvement (over no adaptation) for voice queries compared with long-form data. This observation aligns with the perplexity values shown in the table. The perplexity ($PP$) of the distribution of languages is smaller for voice queries ($PP=33.8$) than for long-form data ($PP=56.2$). If the languages were balanced (equal class priors), the perplexity would be $PP=65$ or the number of languages supported by the system.

The output transform approach has a larger improvement than prior replacement, especially for the long-form domain. This is expected because: (1) the output transform makes use of the LangID model outputs and reference labels on the dev-set, while prior replacement only makes use of the reference labels (related to class priors) of the dev-set; (2) the output transform better compensates for the duration gap between training and inference.

At the same time, it is also worth noting that adaptation with out-of-domain parameters (\emph{i.e.} the shaded area in Table~\ref{tbl:adapt}) will hurt the performance for specific domains. This can be mitigated by using a larger smoothing parameter $R$ in Eq.~\ref{eq:class_prior_prob_estimate} for the prior replacement approach, or using a larger regularization weight $w_\text{reg}$ in Eq.~\ref{eq:adapt_optimize} for the output transform approach.

\section{Conclusion}
\label{sec:conclusion}

In this paper, we described a novel language identification system based on conformer layers and attentive temporal pooling. This model can be parallelized on accelerator hardware, and perform inference in a streaming fashion. Our experiments confirm that conformer based models significantly outperform LSTM or transformer based models under different model size constraints, and are relatively computationally efficient. We also show that attentive temporal pooling further improves performance. We studied two different domain adaptation approaches, namely prior replacement and a discriminatively trained output transform. They allow for the deployment of the same LangID model to different application domains where the prior language distributions and/or data are different, and effectively improve the domain performance.

\section{Acknowledgment}
\label{sec:ack}
The authors thank Benjamin Lee and Jonathan Shen for their help with Lingvo~\cite{shen2019lingvo} integration, and Pedro Moreno Mengibar, Di Li, Bhaskar Gurram, Sunha Ahn, Pavan Desikan and the anonymous Odyssey reviewers for the helpful discussions.

\clearpage
\bibliographystyle{IEEEbib}
\bibliography{Odyssey2022_BibEntries}

\end{document}